\documentclass[aps,prl,twocolumn, showpacs,notitlepage,superscriptaddress]{revtex4-1}

\usepackage{amsmath}
\usepackage{graphicx}
\usepackage{lmodern}
\usepackage{amsmath}
\usepackage{bm}
\usepackage{hyperref}\usepackage{url}
\usepackage{subfigure}%
\usepackage{dsfont}
\usepackage[normalem]{ulem}

\usepackage{amsbsy}
\usepackage{dcolumn}
\usepackage{amsthm}
\usepackage{bm}
\usepackage{esint}
\usepackage{multirow}
\usepackage{hyperref}

\usepackage{cleveref}

\usepackage{mathrsfs}
\usepackage{amsfonts}
\usepackage{amsbsy}
\usepackage{dcolumn}
\usepackage{bm}
\usepackage{multirow}
\usepackage{color}

\usepackage{color}
\usepackage{amssymb}
\newcommand{\beq} {\begin{equation}}
\newcommand{\eeq} {\end{equation}}
\newcommand{\bea} {\begin{eqnarray}}
\newcommand{\eea} {\end{eqnarray}}
\newcommand{\be} {\begin{equation}}
\newcommand{\ee} {\end{equation}}
\renewcommand{\(}{\left(}
\renewcommand{\)}{\right)}
\renewcommand{\[}{\left[}
\renewcommand{\]}{\right]}

\DeclareMathOperator{\sgn}{sgn}
\DeclareMathOperator{\Tr}{Tr}

\begin{document}

\title {Topological Phase Transitions in Multi-component Superconductors}
\author{Yuxuan Wang}
\affiliation{Department of Physics and Institute for Condensed Matter Theory, University of Illinois at Urbana-Champaign, 1110 West Green Street, Urbana, Illinois 61801-3080, USA }
\author{Liang Fu}
\affiliation{Department of Physics, Massachusetts Institute of Technology, Cambridge, Massachusetts 02139, USA}

\begin{abstract}
We study the phase transition between a trivial  and a  time-reversal-invariant topological superconductor {in a single-band system}. By analyzing the interplay of symmetry, topology and energetics, we show that for a generic normal state band structure, the phase transition occurs via extended intermediate phases in which even- and odd-parity pairing components coexist. For inversion-symmetric systems, the coexistence phase spontaneously breaks time-reversal symmetry. For noncentrosymmetric superconductors, the low-temperature intermediate phase is time-reversal breaking, while the high-temperature phase preserves time-reversal symmetry and has topologically protected line nodes. Furthermore, with  approximate rotational invariance, the system has an emergent $U(1)\times U(1)$ symmetry, and novel topological defects, such as half vortex lines binding Majorana fermions, can exist. {We analytically solve for the dispersion of the Majorana fermion and show that it exhibit  small and large velocities at low and high energies.} Relevance of our theory to superconducting pyrochlore oxide Cd$_2$Re$_2$O$_7$ and half-Heusler materials is discussed.
\end{abstract}
\date{\today}

\maketitle

 Topological superconductivity~\cite{Fu-Berg-2010,Kriener-2011,Levy-2013,
Fu-2014,sun-2014,brydon-2014,Wan-2014,Nikosai-2012,
Scheurer-2015,Hosur-2014,Yuan-2014,Yoshida-2015,Yang-2015,
Ando-2015,scheurer-2016,arXiv:1702.03294} offers a unique platform for studying the interplay between topological phases of matter, unconventional superconductivity (SC), and exotic quasiparticle and vortex excitations. In the presence of time-reversal and inversion symmetry, topological superconductors require an odd-parity order parameter (e.g. $p$-wave)~\cite{Fu-Berg-2010,Sato-2010}.
Theoretical studies~\cite{Fu-Berg-2010, Wan-2014} proposed that Cu$_x$Bi$_2$Se$_3$, a doped topological insulator that becomes superconducting below $T_c\sim 3.8$K, has an odd-parity pairing symmetry favored by the strong spin-orbit coupling in its normal state. Recently, a series of experiments including NMR~\cite{Zheng}, specific heat~\cite{Maeno}, magnetoresistance~\cite{Visser,Wen} and torque measurement~\cite{Lu} under a rotating magnetic field have all found that the superconducting state in {Cu-,} Sr-, and Nb-doped Bi$_2$Se$_3$ spontaneously breaks crystal rotational symmetry, only compatible with the time-reversal-invariant $p$-wave pairing with the $E_u$ symmetry~\cite{Fu-Berg-2010, Fu-2014}. There is currently high interest in searching for the topological excitations in these materials~\cite{PhysRevB.86.094507,arXiv:1209.0656,Venderbos-Kozii-Fu, Guinea1,Guinea2,law-2016,1703.02986,PhysRevB.94.180510}.

In this {paper}, we study topological phase transitions in superconductors resulting from the change of pairing symmetry from even- to odd-parity. Our study is motivated by a number of experiments showing that pairing interactions in even- and odd-parity channels are of comparable strength in several materials, hereafter referred to as multi-component superconductors.
In the non-centrosymmetric superconductor Li$_2$(Pd,Pt)$_3$B, the odd-parity spin-triplet and even-parity spin-singlet pairing components vary continuously as a function of the alloy composition~\cite{togano05,sigrist06,zheng12}. In the pyrochlore oxide Cd$_2$Re$_2$O$_7$~\cite{sc_cdreo,hc2}, applying pressure drives phase transitions between different superconducting states, accompanied by an anomalous enhancement of the upper critical field exceeding the Pauli limit~\cite{hc2}. This has been interpreted as a transition from spin-singlet to  spin-triplet dominated superconductivity.
On the theory side, a pairing mechanism for odd-parity superconductivity in spin-orbit-coupled systems has been recently proposed~\cite{Fu-2015, wang2016, ruhman}, where the pairing interaction arises from the fluctuation of an inversion symmetry breaking order. It was found that this interaction is attractive and nearly degenerate~\cite{lederer-kivelson,kang-fernandes,chubukov-15} in the two fully-gapped Cooper channels with $s$-wave and $p$-wave symmetry respectively.

 The topology of a superconductor depends crucially on its order parameter, which is in turn determined by  energetics. Therefore a change of order parameter as a function of tuning parameters and temperature can result in a topological phase transition in multi-component superconductors. Furthermore,
spontaneous time-reversal-symmetry breaking can be energetically favored in the transition region, thus changing the symmetry that underlies the classification of topological superconductors~\cite{ludwig}. Both energetics and spontaneous symmetry breaking need to be taken into account in theory of topological phase transitions in superconductors.


We show that the phase diagram of multi-component superconductors is largely determined by the fermiology of the normal state, rather than the microscopic pairing mechanism (which is often not exactly known). We find two types of phase diagrams for generic Fermi surfaces with and without inversion symmetry, shown in Fig.1 panel (b) and (d).
Remarkably, we find that the transition between the $s$-wave-dominated trivial phase and the $p$-wave-dominated topological phase is {\it generically} interrupted by an extended intermediate phase where $s$-wave and $p$-wave pairings coexist. For superconductors with inversion symmetry, the intermediate phase is a spontaneous time-reversal symmetry breaking (TRSB) and inversion symmetry breaking superconducting state with $s$-wave and $p$-wave order parameters differing by a fixed relative phase of $\pm \pi/2$~\cite{congjun-trsb,maiti2013,Wang-2014,hinojosa2014}. 
This $s+ip$ state realizes a superconducting analog of axion insulator~\cite{Qi-Hughes-Zhang-2008,PhysRevLett.102.146805,PhysRevLett.108.146601,wang-zhang-2013}, and exhibits thermal Hall conductance on the surface. For noncentrosymmetric superconductors~\cite{samokhin1,samokhin2, agterberg-review},  we predict two intermediate phases in the transition region at different temperatures: a time-reversal-invariant phase at temperatures close to $T_c$, and a time-reversal-breaking phase at low temperature. In particular, the time-reversal-invariant phase has topologically protected {\it line} nodes in the bulk~\cite{ken-14,chiu-schnyder-15}.

We derive the above results by general considerations of symmetry, topology and energetics. Important to our analysis is an emergent $U(1)\times U(1)$ symmetry associated with the two phases of $\Delta_{\pm}\equiv \Delta_s\pm \Delta_p$, where $\Delta_{s}$ and $\Delta_p$ are the $s$-wave and $p$-wave superconducting order parameters respectively. In the special case of isotropic Fermi surface, the $U(1)\times U(1)$ symmetry is exact at the transition between $s$-wave and $p$-wave pairing symmetry, and leads to a direct first-order phase transition between trivial and topological superconductors; see Fig.\ 1 panel (a) and (c). In the general case of superconductors with anisotropic Fermi surfaces and gaps, the $U(1)\times U(1)$ symmetry near the topological phase transition is approximate and provides a useful starting point for our theory. {Moreover, in this regime, half quantum vortices, which corresponds to the winding of one of $U(1)$ phases~\cite{babaev}, can appear as topological defects, which bind chiral Majorana modes. {We solve for the dispersion of the Majorana mode, and show it has a small velocity at zero energy and a large velocity near gap edge.}} 

Our theory is largely independent of specific band structures or pairing mechanisms, and is potentially applicable to a broad range of materials. At the end of this work, we discuss the relevance of our general results for the superconducting phases of pyrochlore oxide Cd$_2$Re$_2$O$_7$ and half-Heusler compounds, and make testable predictions.


\begin{figure}
\includegraphics[width=\columnwidth]{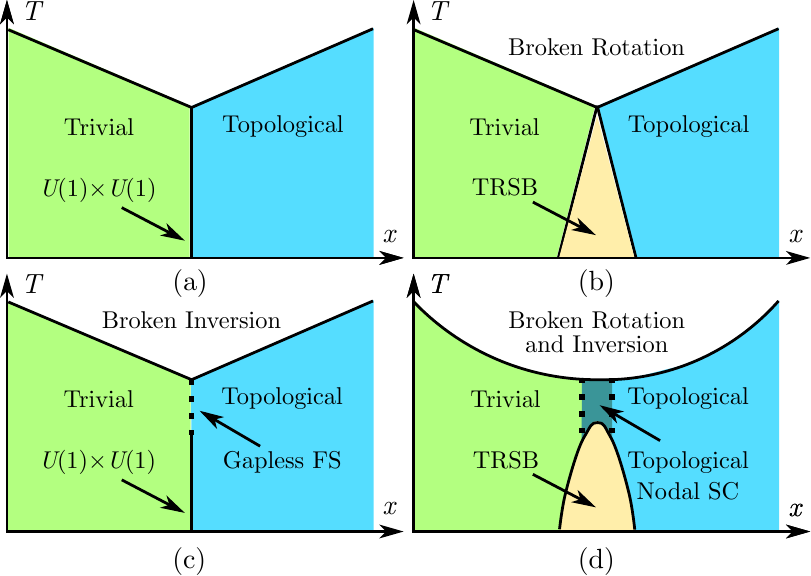}
\caption{Schematic phase diagrams near a topological phase transition in multi-component superconductors with [Panels (a,b)] and without [Panels (c,d)] {inversion} symmetry, and with [Panels (a,c)] and without [Panels (b,d)] {rotational} symmetry. In (a,b), the ``trivial" phase has an $s$-wave pairing symmetry and the ``topological" phase is $p$-wave.  In (c,d) without inversion symmetry, the topological phase corresponds to the region where $p$-wave component is larger. In Panel (c) at the dashed line one of the spin-textured Fermi surface is completely gapless, while in panel (d) in the region between the dashed lines the superconducting states have topologically protected line nodes on the Fermi surface. }
\label{phase}
\end{figure}

 {\it $U(1)\times U(1)$ symmetry.---}  Throughout this work, we assume the system under study has strong spin-orbit coupling. Then single-particle energy eigenstates in the normal state generally do not have well-defined spin. Nonetheless, when both time-reversal and inversion symmetry are present, energy bands remain doubly degenerate at every momentum $\bf k$, which we label with pseudo-spin index $\sigma$. We choose to work in the manifestly covariant Bloch basis \cite{PhysRevLett.115.026401}, where the state $|{\bf k},\sigma=\pm \rangle$ has the same symmetry property as the spin eigenstate $|{\bf k}, s_z =\uparrow (\downarrow)\rangle$ under the joint rotation of electron's momentum and spin.

  As a convenient starting point, we first consider systems with full rotational invariance. In such systems, all the pairing order parameters can be classified by their total ($J$) angular momentum. We focus on $J=0$ pairings with a full gap. If inversion symmetry is present, there are two types of $J=0$ order parameters, with even- or odd-parity respectively. The even-parity $J=0$ pairing has $s$-wave orbital angular momentum given by  $\mathcal{H}_s=\Delta_s c^\dagger_{\bf k}i\sigma^y (c^\dagger_{-\bf k})^T$, while the odd-parity $J=0$ pairing has $p$-wave orbital  angular momentum given by
$\mathcal{H}_p= \Delta_p c^\dagger_{\bf k}(\hat{\bf k}\cdot \vec\sigma)i\sigma^y (c^\dagger_{-\bf k})^T$.
This $p$-wave order parameter looks similar to that of $^3$He-$B$ phase, but the spin quantization axis is rigidly locked to the momentum by spin-orbit coupling here. 
In both 2D and 3D, $\Delta_p$ realizes time-reversal-invariant topological superconductivity in the DIII class.

We now analyze the interplay between $s$-wave and $p$-wave pairings. Generically, the free energy is given by
\begin{align}
\mathcal{F}=&\alpha_1|\Delta_s|^2+\alpha_2|\Delta_p|^2+\beta_1|\Delta_s|^4+\beta_2|\Delta_p|^4\nonumber\\
&+4\bar\beta|\Delta_s|^2|\Delta_p|^2+\tilde\beta(\Delta_s^2\Delta_p^{*2}+\Delta_p^2\Delta_s^{*2}).
\label{Fsp}
\end{align}
The temperature-dependent coefficients $\alpha_1, \alpha_2$ are determined by the microscopic pairing mechanism.
We are interested in the case when $s$-wave and $p$-wave instabilities are comparable in strength, i.e., when $\alpha_1\sim\alpha_2$, 
so that tuning some parameters such as pressure or chemical composition can drive a phase transition. The interplay between $s$- and $p$-wave order parameters is controlled by the $\beta$ coefficients only. It is important to note that, {within weak-coupling theory,} $\beta$'s do not rely on pairing interactions, and are completely determined by the normal state electronic structure, as shown from the Feynman diagram calculation  (for details see \cite{SM}). Explicitly evaluating these diagrams, we obtain that
$\beta_1=\beta_2=\bar\beta=\tilde\beta\equiv \beta = 5\zeta(3)/(8\pi^2T^2N(0))$,
where $N(0)$ is the density of states, and $\zeta(x)$ is the Riemann zeta function.

The last term in \eqref{Fsp} is minimized when  the phase difference of the two order parameters at $\Delta\phi=\pm \pi/2$. Under this condition, at the phase boundary $\alpha_1=\alpha_2=\alpha$, the free energy \eqref{Fsp} becomes
\be
\mathcal{F}=\alpha(|\Delta_s|^2+|\Delta_p|^2)+\beta(|\Delta_s|^2+|\Delta_p|^2)^2.
\ee
This free energy possesses a $U(1)\times U(1)$ symmetry~\cite{wang2016} associated with the common phase and relative amplitude of $\Delta_{s,p}$.~\footnote{Note that he additional $U(1)$ is \emph{not} the rotational symmetry we impose, but rather is {emergent}.}
When $\alpha_1\neq\alpha_2$, the $U(1)\times U(1)$ symmetry is broken, and the free energy is minimized such that the pairing channel with higher transition temperature (i.e. smaller $\alpha$) completely suppresses the other, and the phase transition is of first-order. Thus we obtain the phase diagram shown in Fig.\ \ref{phase}(a). {In a previous work~\cite{GR} it was reported for a rotational invariant system there is a coexistence phase with both $s$-wave and $p$-wave orders. Our results differ here, and as we shall see, to obtain the coexistence phase it is necessary to break the rotational invariance, at least within weak-coupling theory.}

The emergent $U(1)$ symmetry is a general consequence of the rotational and inversion symmetry of the assumed normal state electronic structure.
To see this more explicitly, it is instructive to divide pseoudo-spin degenerate states on the Fermi surface into two groups, with \emph{helicty} $\chi={\vec \sigma}\cdot \hat {\bf k}=\pm 1$ separately.
Then, the $\Delta_s$ and $\Delta_p$ order parameter both correspond to pairing within each group of helicity eigenstates (which we denote by $\Delta_{\pm}$), with constant gap over the Fermi surface as dictated by rotational invariance. The difference of $\Delta_s$ and $\Delta_p$ is that they are even- and odd-combinations of $\Delta_{\pm}$, i.e., $\Delta_{s,p}=(\Delta_{+}\pm\Delta_-)/\sqrt{2}$~\cite{wang2016,PRL109.187003,brydon-2014}.
In terms of $\Delta_\pm$, the generic free energy \eqref{Fsp} can be rewritten as 
 \begin{align}
 \mathcal{F}=&\alpha(|\Delta_+|^2+|\Delta_-|^2)+\delta\alpha(\Delta_+\Delta_-^*+\Delta_+^*\Delta_-)\nonumber\\
 &+\beta(|\Delta_+|^4+|\Delta_-|^4),
 \label{fpm}
 \end{align}
where the coefficients $\alpha,\beta$ for $\Delta_{\pm}$ terms are identical due to inversion symmetry which transforms opposite helicity eigenstates into each other, and $\delta\alpha\equiv (\alpha_1-\alpha_2)/2$. 

Depending on its sign, $\delta\alpha=\delta\alpha(x)$ controls the relative sign of $\Delta_{\pm}$ in the ground state, i.e., whether $s$-wave or $p$-wave order is favored.
In this form the $U(1)\times U(1)$ symmetry is explicit at $\delta\alpha=0$, i.e., the phase boundary of $s$- and $p$-wave orders. The ``second $U(1)$" can be regarded as a gapless Leggett mode~\cite{Leggett-1966} for the relative phase between $\Delta_{\pm}$.

 {\it Time-reversal symmetry breaking phases.---} In an actual system without full rotational invariance, 
 the $U(1)\times U(1)$ symmetry is at best approximate. To see this, we still consider $s$-wave and $p$-wave pairing orders,  $\mathcal{H}_s= \Delta_s f_s({\bf k})\, c^\dagger_{\bf k}i\sigma^y (c^\dagger_{-\bf k})^T,
 \mathcal{H}_p= \Delta_p f_p({\bf k}) \, c^\dagger_{\bf k}( \hat{\bf k}\cdot \vec{\sigma})i\sigma^y (c^\dagger_{-\bf k})^T$,
where the form factors $f_{s,p}({\bf k})$ are positive and even functions of $\bf k$. For weak-coupling superconductivity, $f_{s,p}({\bf k})=f_{s,p}(\hat{\bf k})$.
Since there is no further symmetry requirement restricting them,  in general $f_s(\hat{\bf k})\neq f_p(\hat{\bf k})$.  As a concrete example, we constructed a microscopic model~\cite{SM} (see also~\cite{congjun-trsb2}) with instabilities towards both $s$-wave and $p$-wave orders. 

By computing the $\beta$ coefficients~\cite{SM} in Eq.\ {\eqref{Fsp} for generic form factors, we find $\bar\beta=\tilde\beta$ and $\bar\beta^2<\beta_1\beta_2$.
This indicates a coexistence phase of $s$-wave and $p$-wave orders~\cite{Wang-2014}. Thus}
 the first-order transition with $U(1)\times U(1)$ symmetry expands into an intermediate phase. Since $\Delta_s$ and $\Delta_p$ differs by a phase $\pi/2$, this  $s+ip$ state  spontaneously breaks time-reversal symmetry~\cite{maiti2013,Wang-2014,hinojosa2014}. Such state in three dimensions has unconventional thermal response described by a axion topological field theory~{\cite{qwz,PhysRevLett.102.146805,ryu-moore-ludwig,shiozaki2014,GR,stone2016}}, hence can be called an ``axion superconductor". 


{\it Phase diagram without inversion symmetry.---}
For spin-orbit-coupled materials without inversion symmetry, the Fermi surface is  generally spin-split. With rotational symmetry, each spin-split Fermi surface is isotropic and has a definite helicity $\chi=\pm 1$. The free energy, written in terms of the order parameters $\Delta_\pm$ on each of the helical  Fermi surfaces, takes a general form
$\mathcal{F}=\alpha_+|\Delta_+|^2+\alpha_-|\Delta_-|^2+\delta\alpha(\Delta_+\Delta_-^*+\Delta_+^*\Delta_-)+\beta_+|\Delta_+|^4+\beta_-|\Delta_-|^4$.
At the  phase boundary with $\delta\alpha=0$, the free energy retains an explicit $U(1)\times U(1)$ symmetry. There are two separate transition temperatures, corresponding to the onset of $\Delta_{\pm}$ respectively. Away from the $\delta\alpha=0$ point, the two order parameters are always mixed down to zero temperature once either one of them becomes nonzero. We thus obtain the phase diagram in Fig.\ \ref{phase}(c).
 For a negative (positive) $\delta\alpha$, $\Delta_+$ and $\Delta_-$ take the same (opposite) sign.  Switching to $\Delta_{s,p}$ notation, the phase with $\Delta_\pm$ of opposite signs has the $p$-wave pairing component dominating over the $s$-wave pairing. This phase is adiabatically connected to the $p$-wave-only phase in the presence of inversion symmetry, and hence is topological~\cite{Qi-2010a}.

 \begin{figure}
\includegraphics[width=\columnwidth]{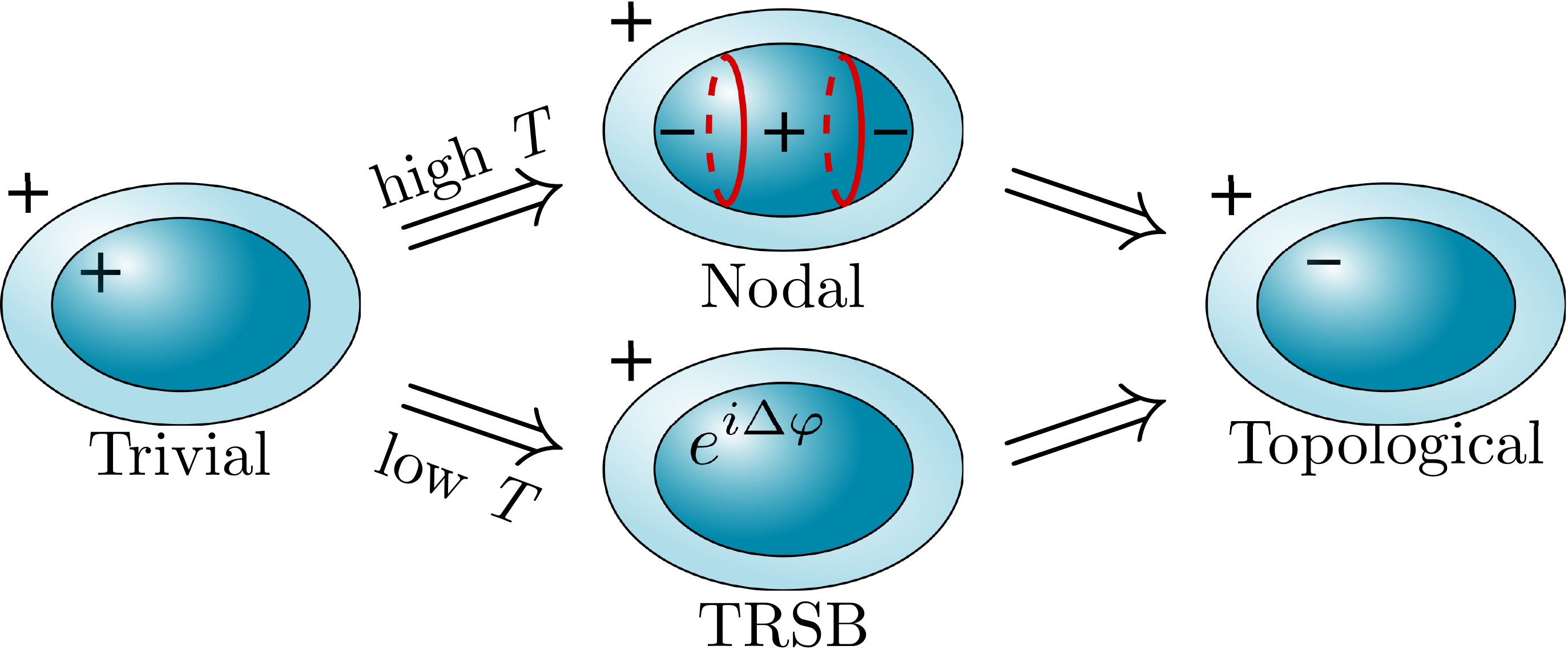}
\caption{Transitions between trivial and topological superconductor with only time reversal symmetry [the case of Fig.\ 1(d)]. In 3D, at high $T$ the transition occurs via intermediate nodal line (nodal points if 2D) superconducting phases, while at low $T$ time-reversal symmetry is spontaneously broken.}
\label{node}
\end{figure}

Finally, {with broken rotational symmetry}, again the low-temperature first-order transition with $U(1)\times U(1)$ symmetry expands into a time-reversal symmetry breaking phase~\cite{SM,chandan2017}, as discussed before. At higher temperatures, $r\equiv \Delta_s/\Delta_p$ is real, and $|r|\gg(\ll)1$ corresponds to a fully-gapped trivial (topological) phase. When $r\sim 1$, the intermediate phase generally have nodes given by $rf_s(\hat{\bf k})= \pm f_p(\hat{\bf k})$, where $\pm$ corresponds to two spin-split Fermi surfaces. It can \emph{only} be satisfied on one of the split Fermi surfaces. The nodes of this intermediate phase have co-dimension 2 and are isolated points in two dimensions and nodal lines in three dimensions. Time-reversal symmetry further requires that in 3D nodal lines appear in pairs and in 2D nodal points in multiples of four (see Fig.\ \ref{node})~\cite{beri-2010}. These nodes are topologically protected by a $\mathbb{Z}_2$ invariant~\cite{ken-14,chiu-schnyder-15}, and lead to flat-bands of surface Andreev states~\cite{flat-band,timm2011,schnyder-ryu}. {The nodal lines are gapped upon entering the time-reversal breaking phase. Time-reversal breaking in nodal line superconductors was obtained in Ref.\ \onlinecite{timm2015}, but only for the surface states; here the time-reversal breaking occurs in the bulk.}
 We summarize the phase diagram in Fig.~\ref{phase}(d).

{\it Experimental consequences.---}
In the time-reversal-breaking phase, e.g. the $s\pm ip$-SC, the surface state can be thought of as a Majorana cone gapped by the $s$-wave component~\cite{SM}. Such a surface state exhibit  thermal Hall effect and polar Kerr effect~\cite{ryu-moore-ludwig,ryu2016}.

\begin{figure}
\includegraphics[width=\columnwidth]{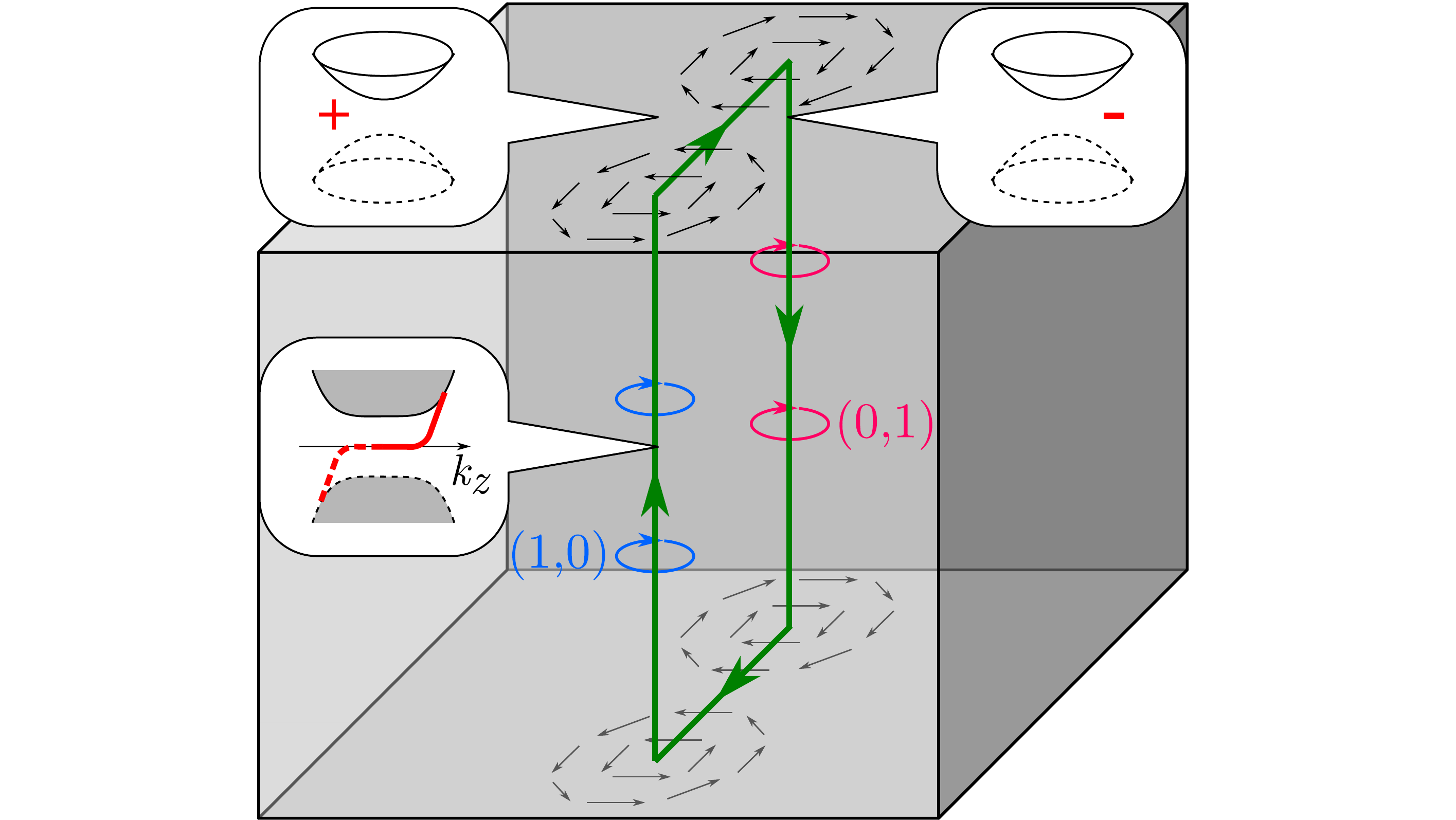}
\caption{The chiral Majorana modes (green arrowed lines) bound to and connecting a pair of half vortices $(1,0)$ and $(0,1)$.  {The surface part of the chiral Majorana mode can be thought of as the chiral edge state at a mass domain wall of the surface Majorana cone. The bulk part of the chiral Majorana mode exhibits a dispersion with both slow and fast modes.}}
\label{hqv}
\end{figure}

When rotational symmetry (even when approximate) is present, half quantum vortices, i.e. the phase winding of only one of $\Delta_{\pm}$ [denoted as $(\pm 1,0)$ and $(0,\pm 1)$], appear as topological defects because of the $U(1)\times U(1)$ symmetry.
The magnetic flux through a half quantum vortex is given by $hc/(4e)$, i.e. half the flux quantum in a superconductor,  hence the name. In 2D, the two helical Fermi surfaces with $\chi=\pm 1$ each enclose a Berry flux of $\pi$, hence their corresponding half quantum vortex for $\Delta_\pm$ binds a \emph{single} Majorana zero mode with non-Abelian statistics~\cite{moore-read, ivanov, Fu-2008}. This is in contrast with a full vortex in a time-reversal-invariant topological superconductor, which binds two Majorana modes with Abelian statistics.
 
 In 3D, 
  the half quantum vortex line binds a propagating chiral Majorana mode~\cite{qwz,stone2016}. 
  Furthermore, we find that the dispersion $\epsilon=\epsilon(k_z)$ of such a chiral Majorana mode exhibit both slow and fast components.
   In \cite{SM} we perturbatively solve the BdG equation for small $k_z\ll \Delta/v_F$, and show that  the dispersion of the chiral Majorana mode is given by $\epsilon(k_z)=v_M k_z$ where
\be
v_M(k_z=0)\approx (\Delta_+/\mu)^2  \log(\mu/\Delta_+) v_F \ll v_F.
\ee
At larger $k_z\sim k_F$, the 2D Fermi surface slice shrinks and the above perturbative result is no longer valid. The vortex mode becomes higher in energy and merges into the bulk with a much larger velocity $v_M\sim v_F$. Therefore the Majorana bound state contains both slow and fast modes, both of which are chiral.
We schematically show such a dispersion in the inset of Fig.\ \ref{hqv}.

Given a pair of opposite half quantum vortices, there exist a pair of chiral Majorana modes on the surface connecting the two vortices. A (0,1) and (1,0) half quantum vortex pair can be viewed as a vortex/antivortex pair for the relative phase $\Delta\varphi=\varphi_+-\varphi_-$ between $\Delta_{\pm}$. Locally, this corresponds to $(1+e^{i\Delta\varphi})s+(1-e^{i\Delta\varphi})p$ symmetry. In the slow-varying spatial limit, across the line where $\Delta\varphi=\pi$, locally the surface states are described by two Majorana cones with opposite mass terms~\cite{SM}, shown in Fig.\ \ref{hqv}. The $\Delta\varphi=\pi$ line acts as a mass domain wall for the Majorana fermions, and thus support a chiral mode. The chiral Majorana modes bound to the half quantum vortices and the surfaces form a closed contour, shown in Fig.\ \ref{hqv}.
  This chiral Majorana mode is charge neutral and can support thermal transport. 
  
 {\it Relation to materials.---} Our theory can be applied to systems where even and odd parity superconducting order parameters are intertwined, such as  Cd$_2$Re$_2$O$_7$ and half-Heusler materials.
 For Cd$_2$Re$_2$O$_7$~\cite{sc_cdreo,hc2}, the anomalous enhancement in upper critical field $H_{c2}$ indicates a symmetry change from spin singlet to spin triplet as a function of pressure. Our theory predicts nodal as well as time-reversal-breaking phases near this region in the phase diagram. In half-Heusler superconductors YPtBi~\cite{paglione1} and LuPtBi~\cite{ylchen}, order parameters with a mixed even- and odd-parity pairings  have been proposed~\cite{agterberg,congjun-3/2} to account for penetration depth measurements~\cite{paglione1}. This microscopic study finds line nodes in a region of mixed-parity phase, consistent with our general phase diagram for noncentrosymmetric superconductors presented in Fig.\ \ref{phase}(d). Our theory further predicts that the superconducting state with line nodes transitions into a new time-reversal breaking phase upon lowering temperatures. 
  It will be interesting to directly search for this time-reversal symmetry low-temperature phase.

 \acknowledgments{This work was supported by the Gordon and Betty Moore Foundation's EPiQS Initiative through Grant No.\ GBMF4305 at the University of Illinois (Y.W.) and the DOE Office of Basic Energy Sciences, Division of Materials Sciences and Engineering under Award No.\ DE-SC0010526 (L.F.).
  We thank the hospitality of the summer program ``Multi-Component and Strongly-Correlated Superconductors" at Nordita, Stockholm, where this work was initiated.
  Y.W. acknowledges support by the 2016 Boulder Summer School for Condensed Matter and Materials Physics through NSF grant DMR-13001648, where part of the work was done.}
%

\renewcommand{\theequation}{S\arabic{equation}}
\renewcommand{\thefigure}{S\arabic{figure}}
\renewcommand{\bibnumfmt}[1]{[S#1]}
\renewcommand{\citenumfont}[1]{S#1}

\onecolumngrid

\newpage
\centerline{\large{\bf Supplemental Material}}

\section{I.~~~Phase diagram from Ginzburg-Landau theory}
\begin{figure}[b]
\includegraphics[width=0.5\columnwidth]{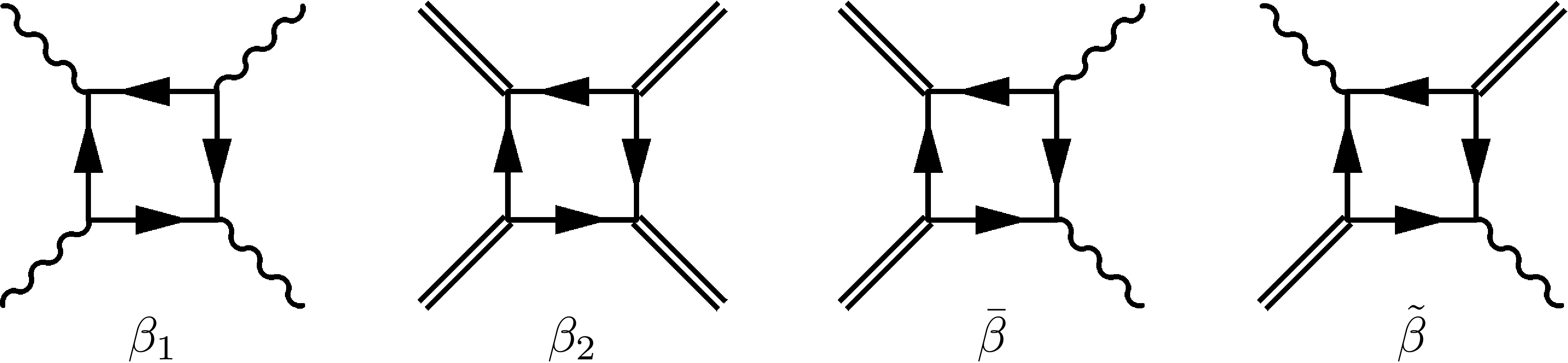}
\caption{The diagrams corresponding to the coefficient $\beta$'s.}
\label{s-beta}
\end{figure}
\subsection{A.~~~With inversion symmetry}

In this Section we present the details on the derivation of the four phase diagrams of Fig.\ 1 of the main text.
First, with inversion symmetry, the Ginzburg-Landau (GL) free energy takes the following form 
\begin{align}
\mathcal{F}=&\alpha_1|\Delta_s|^2+\alpha_2|\Delta_p|^2+\beta_1|\Delta_s|^4+\beta_2|\Delta_p|^4+4\bar\beta|\Delta_s|^2|\Delta_p|^2+\tilde\beta(\Delta_s^2\Delta_p^{*2}+\Delta_p^2\Delta_s^{*2}),
\label{s-Fsps}
\end{align}
and we are interested in regions with $\alpha_1\sim \alpha_2$. The $\beta$ coefficients  are given by evaluating the Feynman diagrams in Fig.\ \ref{s-beta}, where the wavy line and the double line respectively represent $\Delta_s$ and $\Delta_p$. Below we compute these diagrams for cases with and without rotational invariance and obtain the corresponding phase diagrams.

\subsubsection{1.~~~With rotational invariance}
For rotational invariant system, $\Delta_s$ and $\Delta_p$ by symmetry have the same uniform form factors (other than the odd parity part coming from the spin texture for $p$-wave). The $\beta$ coefficients are distinguished by their spin structures and symmetry factors:  
\begin{align}
&\beta_1=\frac{\beta}{2}\Tr[(i\sigma^y)(i\sigma^y)^\dagger(i\sigma^y)(i\sigma^y)^\dagger]=\beta \nonumber\\
&\beta_2=\frac{\beta}{2}\Tr[(i\hat{\bf k}\cdot{\vec \sigma}\sigma^y)(i\hat{\bf k}\cdot{\vec \sigma}\sigma^y)^\dagger(i\hat{\bf k}\cdot{\vec \sigma}\sigma^y)(i\hat{\bf k}\cdot{\vec \sigma}\sigma^y)^\dagger]=\beta \nonumber\\
&\bar\beta=\beta\Tr[(i\hat{\bf k}\cdot{\vec \sigma}\sigma^y)(i\hat{\bf k}\cdot{\vec \sigma}\sigma^y)^\dagger(i\sigma^y)(i\sigma^y)^\dagger] =\beta\nonumber\\
&\tilde\beta=\frac{\beta}{2}\Tr[(i\hat{\bf k}\cdot{\vec \sigma}\sigma^y)(i\sigma^y)^\dagger(i\hat{\bf k}\cdot{\vec \sigma}\sigma^y)(i\sigma^y)^\dagger]=\beta,
\label{s-Fsps2}
\end{align} 
where 
\begin{align}
\beta=\frac{T}{2}\sum_{m}\int\frac{d{\bf k}}{8\pi^3}G^2(\omega_m,{\bf k})G^2(-\omega_m,-{\bf k})=\frac{N(0)T}2\sum_{m}\int\frac{d\epsilon_{\bf k}}{(\omega_m^2+\epsilon_{\bf k}^2)^2}= \frac{5\zeta(3)}{8\pi^2T^2}N(0).
\label{s-eq:beta}
\end{align}
By arguments we have elaborated in the main text, for these values of $\beta$'s the GL free energy has a $U(1)\times U(1)$ symmetry when $\alpha_1=\alpha_2$. When $\alpha_1\neq\alpha_2$ the $U(1)\times U(1)$ symmetry is lifted and ground state is either $s$-wave or $p$-wave. This is shown in Fig.\ 1(a) of the main text.

\subsubsection{2.~~~Without rotational invariance}
Without rotational invariance, the $s$- and $p$-wave order parameters $\Delta_{s,p}$ are generally of different form factors. Namely,
\begin{align}
\mathcal{H}_s=& \Delta_s f_s({\bf k}) \times c^\dagger_{\bf k}i\sigma^y (c^\dagger_{-\bf k})^T, \nonumber\\
 \mathcal{H}_p=& \Delta_p f_p({\bf k}) \times c^\dagger_{\bf k}({\bf k\cdot \sigma})i\sigma^y (c^\dagger_{-\bf k})^T.
 \end{align}
With regards to the $\beta$ coefficients, the Pauli matrix algebra given in \eqref{s-Fsps2} still holds. The only difference here is that the form factors $f_s({\bf k})$ and $f_p({\bf k})$ also enters the integral. Since the dominant contribution comes from near the FS, we can safely set $|{\bf k}|=k_F$ in $f_{s,p}$ and reduce them to angular functions $f_{s,p}(\theta,\phi)$. {Here $(\theta,\phi)$ are angular coordinates on a 3D Fermi surface, and our analysis below extends straightforwardly to 2D cases.} The $\beta$ coefficients are thus given by
\begin{align}
\beta_1&=N(0)T\sum_{n}\int \frac{d\sin\theta d\phi}{4\pi} f_s^4(\theta,\phi)\int\frac{d\epsilon}{(\omega_n^2+\epsilon^2)^2}= \beta\int\frac{d\sin\theta d\phi}{4\pi} f_s^4(\theta)\nonumber\\
\beta_2&=N(0)T\sum_{n}\int \frac{d\sin\theta d\phi}{4\pi} f_p^4(\theta,\phi)\int\frac{d\epsilon}{(\omega_n^2+\epsilon^2)^2}= \beta\int\frac{d\sin\theta d\phi}{4\pi} f_p^4(\theta)\nonumber\\
\bar\beta=\tilde\beta&=N(0)T\sum_{n}\int \frac{d\sin\theta d\phi}{4\pi} f_s^2(\theta,\phi)f_p^2(\theta,\phi)\int\frac{d\epsilon}{(\omega_n^2+\epsilon^2)^2}= \beta\int\frac{d\sin\theta d\phi}{4\pi} f_s^2(\theta,\phi)f_p^2(\theta).
\label{s-eq:beta2}
\end{align}
By expanding $f_s^2$ and $f_p^2$ into spherical harmonics components $Y_{\ell m}(\theta,\phi)$, we have  $\beta_1=\beta\int f_s^4(\theta,\phi) d\sin\theta d\phi/(4\pi)=\beta\sum_{\ell m} S_{\ell m}^2$, $\beta_2=\beta\int f_p^4(\theta,\phi) d\sin\theta d\phi/(4\pi)=\beta\sum_{\ell m} P_{\ell m}^2$, and $\bar\beta=\tilde\beta=\int f_s^2(\theta,\phi)f_p^2(\theta,\phi) d\sin\theta d\phi/(4\pi)=\beta\sum_{\ell m} S_{\ell m}P_{\ell m}$. It is straightforward to verify that 
\be\bar\beta^2=\tilde\beta^2<\beta_1\beta_2,\ee
In the free energy \eqref{s-Fsps}, the minimization of the term $\Delta_s^2\Delta_p^{*2}$ dictates that the relative phase between  $\Delta_s$ and $\Delta_p$ is $\Delta\phi=\pm \pi/2$, and thus
$\Delta_s^2\Delta_p^{*2}=-|\Delta_s|^2|\Delta_p|^2$. After substituting this relation and $\bar\beta=\tilde\beta$ into \eqref{s-Fsps}, we obtain
\begin{align}
\mathcal{F}=\alpha_1|\Delta_s|^2+\alpha_2|\Delta_p|^2+\beta_1|\Delta_s|^4+\beta_2|\Delta_p|^4+2\bar\beta|\Delta_s|^2|\Delta_p|^2. 
\end{align}
Since $\bar\beta^2=\tilde\beta^2<\beta_1\beta_2$, we find that $s$-wave and $p$-wave orders compete but coexist at their phase boundary. In the coexisting phase the time-reversal symmetry (TRS) is broken by the choice of $\Delta\phi=\pm \pi/2$. We show the phase diagram in Fig.\ 1(b) of the main text.

\subsection{B.~~~Without inversion symmetry}
\subsubsection{1.~~~With rotational invariance}
When inversion symmetry is broken, the FS splits into two. However when rotational invariance is intact, one can still define $\Delta_{\pm}$ on each of the split FS. The two pairing fields are coupled to FSs with orthogonal helicity $\chi=\pm 1$, and thus decouple in the free energy except at quadratic level, which can be induced by pair hopping interaction between the two FS's. The free energy is thus
\begin{align}
 \mathcal{F}=\alpha_+|\Delta_+|^2+\alpha_-|\Delta_-|^2+\delta\alpha(\Delta_+\Delta_-^*+\Delta_+^*\Delta_-)+\beta_+|\Delta_+|^4+\beta_-|\Delta_-|^4.
 \label{s-inversion}
\end{align}
This is Eq.\ (7) of the main text. Note that this form of free energy can also be obtained using $\Delta_{s,p}=(\Delta_++\Delta_-)/\sqrt{2}$ and starting from the following form
\begin{align}
\mathcal{F}=&\alpha_1|\Delta_s|^2+\alpha_2|\Delta_p|^2+\beta_1|\Delta_s|^4+\beta_2|\Delta_p|^4+4\bar\beta|\Delta_s|^2|\Delta_p|^2+\tilde\beta(\Delta_s^2\Delta_p^{*2}+\Delta_p^2\Delta_s^{*2})\nonumber\\
&+\alpha'(\Delta_s\Delta_p^*+\Delta_s^*\Delta_p)+\beta'(|\Delta_s|^2+|\Delta_p|^2)(\Delta_s\Delta_p^*+\Delta_s^*\Delta_p)+\bar\beta'(|\Delta_s|^2-|\Delta_p|^2)(\Delta_s\Delta_p^*+\Delta_s^*\Delta_p).
\end{align}
Compared with the inversion symmetric case \eqref{s-Fsps}, terms with linear coupling $\sim\Delta_s\Delta_p^*+\Delta_s^*\Delta_p$ is allowed due to the broken inversion symmetry. To evaluate the coefficients one still evaluates the square diagrams, and for the fermionic Green's function we use
\begin{align}
\hat G(\omega_m,{\bf k})=&(i\omega_m-\epsilon_{\bf k}-\lambda\hat{\bf k}\cdot{\vec \sigma})^{-1}=\frac{i\omega_m-\epsilon_{\bf k}+\lambda\hat{\bf k}\cdot{\vec \sigma}}{[i\omega_m-\epsilon_{\bf k}]^2-\lambda^2},
\end{align}
where $\lambda$ characterizes the splitting of the FS. The derivation has been performed in Ref.\ \onlinecite{s-wang2016} and the resulting free energy is identical to \eqref{s-inversion}. The calculation is rather lengthy and we do not repeat here.

This free energy at $\delta\alpha=0$ has an enhanced $U(1)\times U(1)$ symmetry, and otherwise $\Delta_+$ and $\Delta_-$ are either of the same sign (trivial) or opposite signs (topological) depending on the sign of $\delta\alpha$. The corresponding phase diagram is in Fig.\ 1(c) of the main text.

\subsubsection{2.~~~Without rotational invariance}
In the absence of both rotational invariance and inversion invariance, the free energy has the most general form,
\begin{align}
\mathcal{F}=&\alpha_1|\Delta_s|^2+\alpha_2|\Delta_p|^2+\beta_1|\Delta_s|^4+\beta_2|\Delta_p|^4+4\bar\beta|\Delta_s|^2|\Delta_p|^2+\tilde\beta(\Delta_s^2\Delta_p^{*2}+\Delta_p^2\Delta_s^{*2})\nonumber\\
&+\alpha'(\Delta_s\Delta_p^*+\Delta_s^*\Delta_p)+\beta'(|\Delta_s|^2+|\Delta_p|^2)(\Delta_s\Delta_p^*+\Delta_s^*\Delta_p)+\bar\beta'(|\Delta_s|^2-|\Delta_p|^2)(\Delta_s\Delta_p^*+\Delta_s^*\Delta_p)
\end{align}
where $\beta_1\beta_2>\bar\beta^2=\tilde\beta^2$, due to identical arguments as the case with inversion symmetry (it turns out the splitting of the FS does not affect the relation~\cite{s-wang2016}). To simplify the calculation we rescale $\Delta_p$ and all the coefficients such that $\beta_1=\beta_2\equiv \beta_0$, and then we have $\bar\beta=\tilde\beta<\beta_1=\beta_2$. We define  $\beta''\equiv \beta_0-\bar\beta>0$, $\alpha=(\alpha_1+\alpha_2)/2$ and $\delta\alpha=(\alpha_1-\alpha_2)/2$. After doing this, we can rewrite the free energy as
\begin{align}
\mathcal{F}=&\alpha(|\Delta_s|^2+|\Delta_p|^2)+\delta\alpha(|\Delta_s|^2-|\Delta_p|^2)+\beta_0(|\Delta_s|^2+|\Delta_p|^2)^2+(\beta_0-\beta'')(\Delta_s\Delta_p^*+\Delta_s^*\Delta_p)^2 - 2\beta''|\Delta_s|^2|\Delta_p|^2\nonumber\\
&+\alpha'(\Delta_s\Delta_p^*+\Delta_s^*\Delta_p)+\beta'(|\Delta_s|^2+|\Delta_p|^2)(\Delta_s\Delta_p^*+\Delta_s^*\Delta_p)+\bar\beta'(|\Delta_s|^2-|\Delta_p|^2)(\Delta_s\Delta_p^*+\Delta_s^*\Delta_p).
\label{s-FFsp}
\end{align}
To connect with previous cases, when inversion symmetry is intact, we have $\alpha'=0$, $\beta'=0$, and $\bar\beta'=0$, and when rotational invariance is intact, we have $\beta''=0$ and $\bar\beta'=0$. 
 
Now we analyze the ground states given by this free energy. Note that in \eqref{s-FFsp}, there exist two types of phase coupling terms, namely $\Delta_s\Delta_p^*+\Delta_s^*\Delta_p$ and $(\Delta_s\Delta_p^*+\Delta_s^*\Delta_p)^2$. The interplay between these terms can drive a time-reversal symmetry breaking transition inside the SC phase, as we argued in the main text. However, the direct minimization of this free energy is rather lengthy. 

The physics observed above is more transparent in an alternative approach. To this end, we define auxiliary order parameters
\be
\Delta_1=(\Delta_s+\Delta_p)/\sqrt2,~~~\Delta_2=(\Delta_s-\Delta_p)/\sqrt2.
\ee
Note that in the absence of rotational symmetry, $\Delta_{1,2}$ are {\it not} uniform SC gaps on helical FS's with $\chi=\pm 1$ (namely $\Delta_{\pm}$), since $\Delta_{s,p}$ have distinct form factors. In terms of $\Delta_{1,2}$, the free energy becomes
\begin{align}
\mathcal{F}=&(\alpha+\alpha')|\Delta_1|^2+(\alpha-\alpha')|\Delta_2|^2+ \delta\alpha (\Delta_1\Delta_2^*+\Delta_1^*\Delta_2)\nonumber\\
&+(2\beta_0+\beta'-3\beta'')|\Delta_1|^4+(2\beta_0-\beta'-3\beta'')|\Delta_2|^4 \nonumber\\
&+2\beta''|\Delta_1|^2|\Delta_2|^2+\bar\beta'(|\Delta_1|^2-|\Delta_2|^2)(\Delta_1\Delta_2^*+\Delta_1^*\Delta_2)+2\beta''(\Delta_1^2\Delta_2^{*2}+\Delta_1^{*2}\Delta_2^2).
\label{s-freeE}
\end{align}

Recall that $\delta\alpha(\Delta_1\Delta_2^*+\Delta_1^*\Delta_2)\equiv \delta\alpha(|\Delta_s|^2-|\Delta_p|^2)$ term distinguishes the onset temperature for $s$-wave (trivial) and $p$-wave (topological) SC orders, where $\delta\alpha=\delta\alpha(x)$ is tuned by a parameter $x$. To be specific, we assume that $\delta\alpha(x)=x-x_0$. On the other hand, upon lowering the temperature, $\alpha$ becomes negative, which drives the SC transitions. To be specific, we assume that $\alpha(T)=(T-T_0)$ (for our purposes the units are unimportant). Without loss of generality, we also assume $\alpha'<0$, which ensures that $\Delta_1$ component dominates.

We now compute the phase diagram in terms of $x$ and temperature $T$. Upon lowering temperature, superconductivity sets in when one of the eigenvalues of the quadratic form in Eq.\ \eqref{s-freeE} first changes sign, i.e., when $\alpha-\sqrt{\alpha'^2+\delta\alpha^2}=0$. Thus the critical temperature of SC is
\begin{align}
T_c(x)=T_0+\sqrt{(x-x_0)^2+\alpha'^2}.
\end{align}

Below this temperature, in general both $\Delta_1$ and $\Delta_2$ are nonzero in the ground state, since the two are linearly coupled. In this situation, for a generic $x$ and $T$, calculating the ground state is usually a tedious task. To capture the essential physics (i.e., to search for a second transition) with relatively simple calculations, we focus on a special line in the phase space with
\be 
\delta\alpha=-\bar\beta'|\Delta_1|^2,~~~\Delta_2=0.
\label{s-node}
\ee
 Along this line, the second transition is characterized by when $\Delta_2$ acquires a nonzero expectation value. 

 One can easily verify that these two conditions are consistent, at least right below $T_c$  -- substituting \eqref{s-node} into \eqref{s-freeE}, the linear coupling between $\Delta_1$ and $\Delta_2$ vanishes, and right below $T_c$ indeed $\Delta_2=0$ is a local minimum of the \eqref{s-freeE}. In this case only $\Delta_1$ is nonzero, and by simple math
 \begin{align}
 |\Delta_1|=\sqrt{-\frac{\alpha+\alpha'}{2(2\beta_0+\beta'-3\beta'')}}.
 \label{s-17}
 \end{align} 
 Substituting \eqref{s-17} back to \eqref{s-node}, we find that this special line for which $\Delta_2$ vanishes below $T_c$ is given by
 \be
 x=x_0+\frac{\bar\beta'(T-T_0+\alpha')}{2(2\beta_0+\beta'-3\beta'')},
 \label{s-line}
 \ee
 i.e., a straight line in $x$-$T$ phase diagram which intersects with $T_c(x)$ at $x=x_0$.
 
When the temperature further lowers, the quadratic coefficient for $\Delta_2$ also gets negative and ultimately $\Delta_2$ becomes nonzero. When this happens, the minimization of the last term of \eqref{s-freeE} ``locks" the relative phase between $\Delta_1$ and $\Delta_2$ at $\Delta\varphi=\pm\pi/2$, the choice of which breaks an additional time-reversal symmetry. In terms of the original order parameters $\Delta_{s,p}=(\Delta_1\pm \Delta_2)/\sqrt{2}$, we have that $|\Delta_s|=|\Delta_p|$ but $\Delta_{s,p}$ differ by a fixed phase $\phi_0=2\arctan (\Delta_2/\Delta_1)$, i.e., this is a $s+e^{i\phi_0} p$ state.

 This temperature is given by $\alpha-\alpha'-2\beta''|\Delta_1|^2=0$, which results in
\be
T_{\rm TRSB}^0=T_0+\frac{2\beta_0+\beta'-4\beta''}{2\beta_0+\beta'-2\beta''}\alpha',~~~(\alpha'<0),
\ee
as a secondary transition temperature within the SC state.

Away from this line given by \eqref{s-line}, the situation is more complicated, in the sense that right below $T_c(x)$ the two order parameters $\Delta_{1,2}$ are always mixed. Nevertheless, by continuation, a TRSB transition still exists. The only difference is that the presence of both $\Delta_1\Delta_2^*+\Delta_1^*\Delta_2$ and $\Delta_1^2\Delta_2^{*2}+\Delta_1^{*2}\Delta_2^2$ terms dictates that at small $\Delta_{1,2}$ the relative phase is locked at $\Delta\phi=0$ (for negative $\delta\alpha$) or $\Delta\phi=\pi$ (for positive $\delta\alpha$), and only deeper into the SC phase the quartic coupling term becomes important and sets the relative phase at a nontrivial TRSB value $\Delta\varphi=\pm \varphi_0$. It is straightforward to verify that in terms of $\Delta_{s,p}$, this corresponds to a $s+ae^{i\phi_0}p$ state, where $a$ is a real constant.  Due to the presence of linear coupling terms which tends to perserve TRS, the TRSB temperature is generally lower than $T_{\rm TRSB}^0$.

\begin{figure}
\includegraphics[width=0.35\columnwidth]{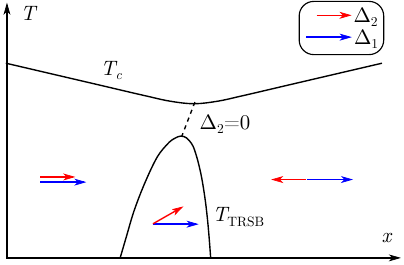}
\caption{The phase diagram in the case without rotational invariance and inversion symmetry. The red and blue arrows denotes the relative phases of $\Delta_1$ and $\Delta_2$.}
\label{s-phasesm}
\end{figure}
With these results we can plot the global phase diagram, as shown in Fig.\ \ref{s-phasesm}. Note that compared with the case with inversion symmetry, the TRSB phase gets detached from the $T_c(x)$ line. This is reminiscent of the phase diagram in Ref. \onlinecite{s-maiti2013}.
Also note that when $\Delta_2=0$, $\Delta_s=\Delta_p$. Since $\Delta_s$ and $\Delta_p$ in the absence of rotational symmetry have different form factors (see next section for more details), the resulting SC state becomes nodal. This is in contrast with the rotational invariant case, where $\Delta_s=\Delta_p$ leaves one of the helical FS fully gapless. In this case, even when $\Delta_s\neq\Delta_p$ but remains close, the SC node survives in a finite but small region of the phase diagram. As we discussed in the main text, the nodes are located on one of the FS's,  and form isolated points in 2D and nodal lines in 3D. This phase diagram in Fig.\ \ref{s-phasesm} with a nodal region is shown schematically in Fig.\ 1(d) in the main text.

\section{II.~~~A microscopic model for time-reversal symmetry breaking phase}
In the main text and in the previous section, we have used that fact that without rotational invariance, the $s$- and $p$-wave form factors are in general different and the resulting GL coefficient relation ensures a phase with broken time-reversal symmetry. In this section, we explicitly construct a simple 2D model to compute the $s$- and $p$-wave form factors.

\begin{figure}[h]
\includegraphics[width=0.2\columnwidth]{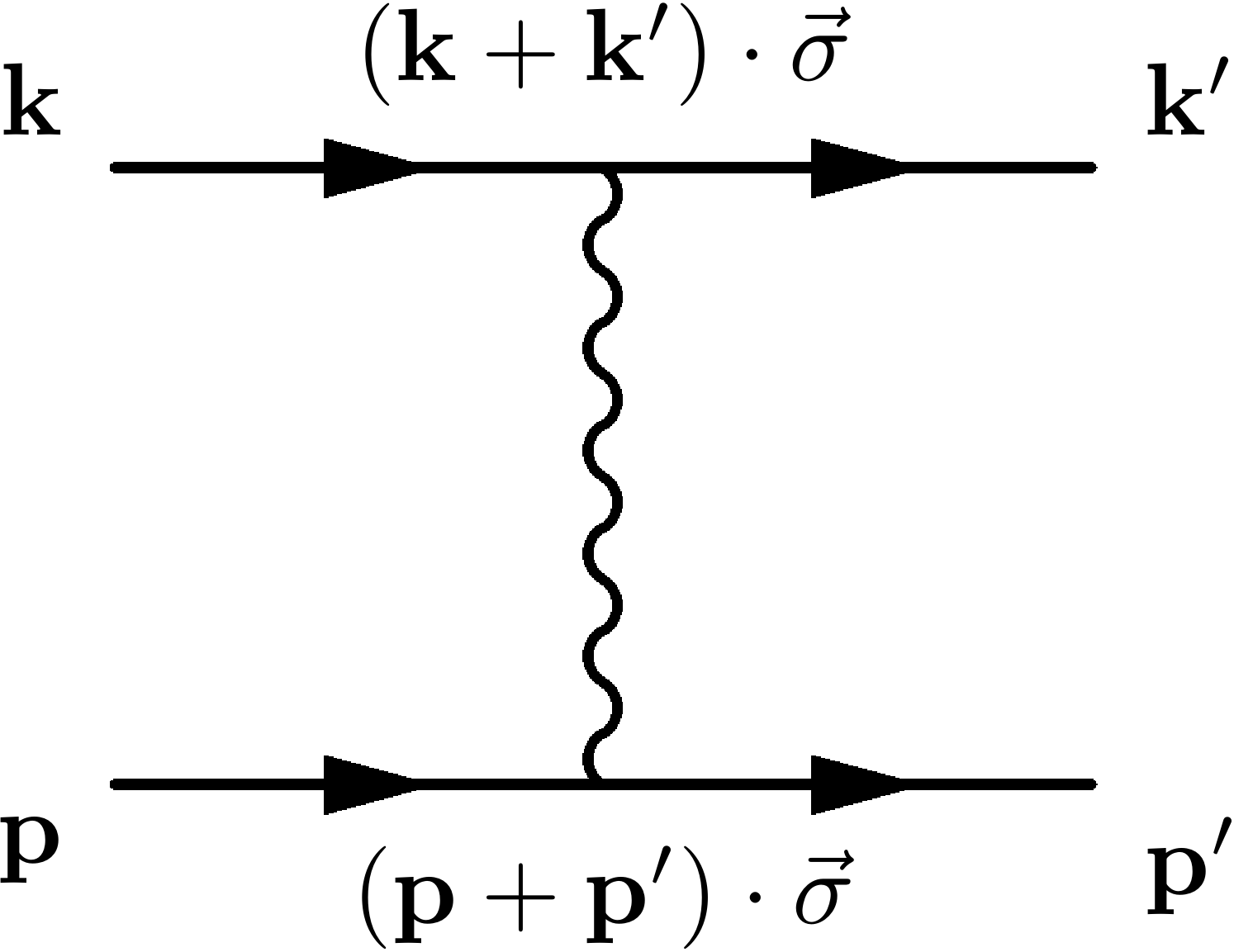}
\caption{The effective fermion-fermion interaction mediated by parity fluctuations.}
\label{s-U}
\end{figure}
 In this model, fermions form a spin-degenerate  elliptical  Fermi surface, and superconductivity is mainly driven by parity fluctuations as proposed in Refs. \onlinecite{s-Fu-2015, wang2016}, namely,
\begin{align}
U_{\alpha\beta,\gamma\delta}({\bf k},{\bf k'}, {\bf p}, {\bf p'})=V [(\hat {\bf k}+\hat {\bf k'})\cdot \vec\sigma_{\alpha\beta}] [(\hat {\bf p}+\hat{\bf p'})\cdot \vec\sigma_{\gamma\delta}]
\end{align}
which is shown diagrammatically in Fig.\ \ref{s-U}. The specific form of the vertex function of this pairing interaction decouples fermions with helicity $\chi=\pm 1$, therefore the pairing gap on the two helical FS's can either be of the same sign ($s$-wave) or with opposite signs ($p$-wave). To lift this degeneracy, we add weak interactions that favors either order. The relative strength of the $p$-wave favoring and $s$-wave favoring interactions acts as the tuning parameter $x$ on the horizontal axis of the phase diagrams shown in the main text. Specifically, we include a phonon-mediated interaction to favor the $s$-wave order, and another interaction mediated by ferromagnetic fluctuations~\cite{s-Leggett-1975,Vollhardt-1990} to favor the $p$-wave order. The effective interaction mediated by ferromagnetic fluctuation (with magnetic moment $\sigma^z$) is given by
\begin{align}
U^{\rm sf}_{\alpha\beta,\gamma\delta}({\bf k},{\bf k'}, {\bf p}, {\bf p'})=V_{\rm sf}(\theta-\theta')  \sigma^z_{\alpha\beta} \sigma^z_{\gamma\delta}.
\end{align}
{Here $\theta$ and $\theta'$ are Fermi surface angles, and $V_{\rm sf}(\theta-\theta')$ is the (static) correlation function of the spin fluctuations projected to the Fermi surface.}
 Admittedly this is only an artificial model, but again, our purpose here is to exemplify a general conclusion.

We note that the weak interactions favoring $s$-wave and $p$-wave orders generally have different preferences for the SC form factors on a given helical FS. This is due to their different dependences on momentum transfer. In particular, the ferromagnetic fluctuations typically are sharply peaked at zero momentum transfer, thus the form factor of the $p$-wave order, which it enhances, tends to be more concentrated on FS regions with highest local density of states to maximize condensation energy. On the other hand, since the spin-fluctuation mediated interaction is repulsive in the $s$-wave channel, the form factor of the $s$-wave order becomes more concentrated on FS regions with lowest local density of states to avoid the repulsion.

To see this more explicitly, we write down the linear SC gap equations (at $T=T_c$) for the two helical FS's,
\begin{align}
\Delta_+(\theta)=&\log\frac{\Lambda}T_c\int d\theta' N(\theta')\[V\cos^2\(\frac{\theta-\theta'}{2}\)\Delta_++V_{\rm ph}(\Delta_+ + \Delta_-) -V_{\rm sf}(\theta-\theta')\Delta_-\]\nonumber\\
\Delta_-(\theta)=&\log\frac{\Lambda}T_c\int d\theta' N(\theta')\[V\cos^2\(\frac{\theta-\theta'}{2}\)\Delta_-+V_{\rm ph}(\Delta_+ + \Delta_-) -V_{\rm sf}(\theta-\theta')\Delta_+\],
\end{align}
where we have assumed weak-coupling pairing and we have integrated over frequency and the direction transverse to the FS. The factor $\cos^2[(\theta-\theta')/2]$ comes from the inner product of spinors aligned with $\bf k$ and $\bf k'$~\cite{s-wang2016}, and in relating $\Delta_+$ with $\Delta_-$ for the $V_{\rm sf}$ term we have used the fact that $\sigma_z\chi_+(i\sigma^y)\chi_+^T\sigma_z=-\chi_-(i\sigma^y)\chi_-^T$ where $\chi_{\pm}=(1\pm\vec\sigma\cdot {\bf k})/2$. For an elliptical FS, the local density of states $N(\theta)\propto 1/|v_F(\theta)|$ around the FS is not a constant and is peaked at the long-axis direction. A simple calculation for dispersion $E=k_x^2/(2m_1) + k_y^2/(2m_2)$ shows
\begin{align}
N(\theta)=\frac{N_0m_1m_2}{m_2\cos^2\theta+m_1\sin^2\theta}
\end{align}

\begin{figure}
\includegraphics[width=0.5\columnwidth]{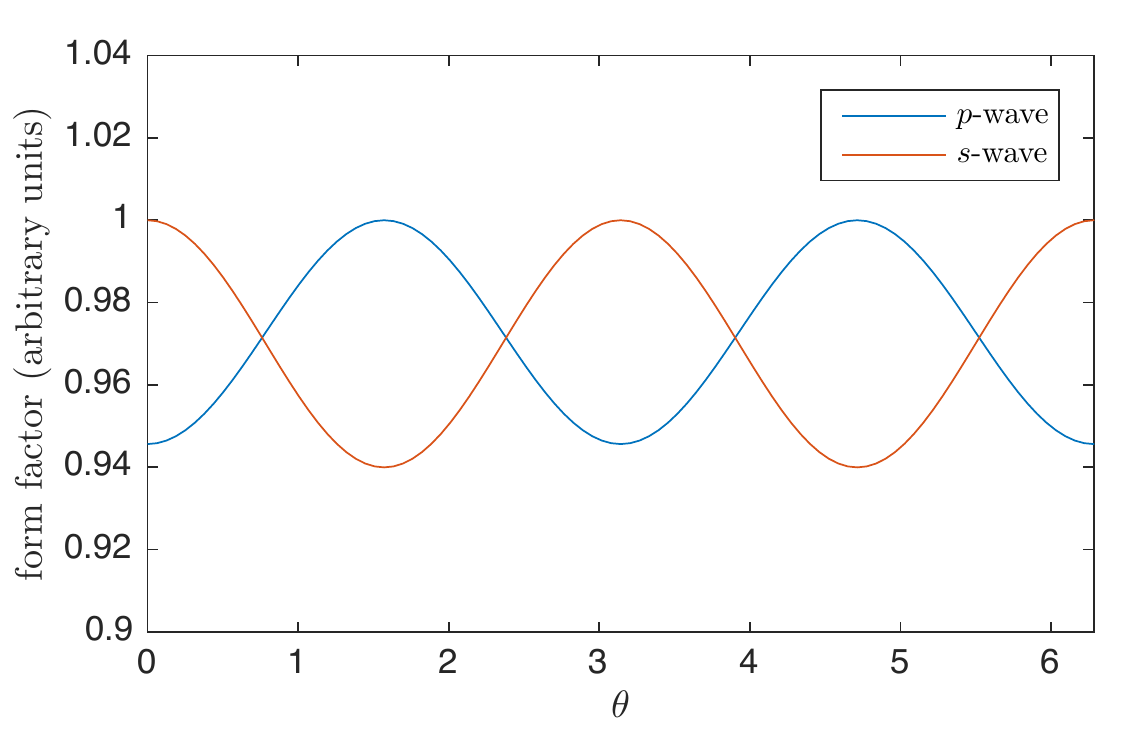}
\caption{The form factors for leading $s$-wave and $p$-wave orders for the helical Fermi surface $\chi=1$. The $p$-wave form factor on the $\chi=-1$ FS is negative by odd parity.}
\label{s-form}
\end{figure}

This set of linear gap equations can be easily solved numerically, say, by modeling $V_{\rm sf}=V_{\rm sf}{^{(0)}} \cos^4(\theta-\theta')$, and we show the form factors of the leading instabilities (i.e. with largest eigenvalue) are indeed different on a given helical FS -- that $s$-wave form factor is peaked around $\theta=0,\pi$ and the $p$-wave form factor is peaked around $\theta=\pm \pi/2$. We plot the form factors of $s$-wave and $p$-wave orders for a given helical FS with $\chi=1$ in Fig.\ \ref{s-form}.

It is straightforward to verify that for these form factors, we indeed have $\beta_1^2+\beta_2^2>2\bar\beta=2\tilde\beta$. Applying our general criterion obtained in last Section, we indeed have a TRSB phase with $s+ip$ order near the phase boundary of $s$-wave and $p$-wave orders, i.e., when the strength of the phonon coupling and the spin-fluctuation coupling are tuned to be comparable.

\section{III.~~~Chiral Majorana modes at the half quantum vortex cores }
In this section we derive the chiral Weyl-Majorana modes bound to the core of a half quantum vortex line in 3D. In 3D, the helicity operator $\chi=\vec\sigma\cdot \mathbf{k}$ is simply the {\it chirality} of a Weyl Fermi surface.
A half quantum vortex is a topological defect around which only one chiral pairing field winds by $2\pi$, which is commonly denoted as $(0,\pm 1)$ or $(\pm1,0)$. In this situation, to compute the vortex line energy spectrum it suffices to only consider the Weyl FS with a winding pairing field, since all states from the other Weyl FS are gapped.

Without loss of generality, we consider the following BdG Hamiltonian
\begin{align}
\mathcal{H}({\bf k})=\(\begin{array}{cc}
\vec\sigma\cdot {\bf k} -\mu & \Delta e^{i\theta} \\
\Delta e^{-i\theta} & -\vec\sigma\cdot {\bf k} +\mu
\end{array}
\)
\label{s-hqv}
\end{align}
where $\theta$ is the real space polar angle in the $xy$ plane. Such a Hamiltonian describes a half vortex line (1,0) in $z$ direction. We look for the dispersion $E=E(k_z)$ of the  in-gap modes bound to the vortex line. Our strategy will be to first solve for problem at $k_z=0$, which is a Majorana bound state, and then treat small $k_z$ as a perturbation and obtain the $k_z$ dispersion.

It is instructive to first analyze the symmetry of Hamiltonian \eqref{s-hqv}, which strongly restricts the from of the low-energy wave function. First, there is a particle-hole symmetry, given by $\mathcal{C}=\sigma^y\tau^y$ ($\tau^y$ is the Pauli matrix in Nambu space), such that
\begin{align}
\mathcal{C}\mathcal{H}^T(-{\bf k})\mathcal{C}^{-1}= -\mathcal{H}({\bf k}).
\end{align}
Second, note that \eqref{s-hqv} is invariant under $\theta\to \theta+\alpha$, $\sigma^{\pm}\to \sigma^{\pm} e^{\pm i\alpha}$, and $\tau^{\pm}\to \tau^{\pm} e^{\mp i\alpha}$, where $\sigma^{\pm}=\sigma^{x}\pm i \sigma^y$. The transformation of $\sigma^{\pm}$ is dictated by the $\vec\sigma\cdot {\bf k}$ coupling, and that of $\tau^{\pm}$ can be verified explicitly. This rotational invariance indicates that one can define a conserved angular momentum \begin{align}
j_z=\ell_z+\sigma_z/2-\tau_z/2.
\end{align}
We expect the in-gap states to have $j_z=0$.
\begin{figure}
\includegraphics[width=0.5\columnwidth]{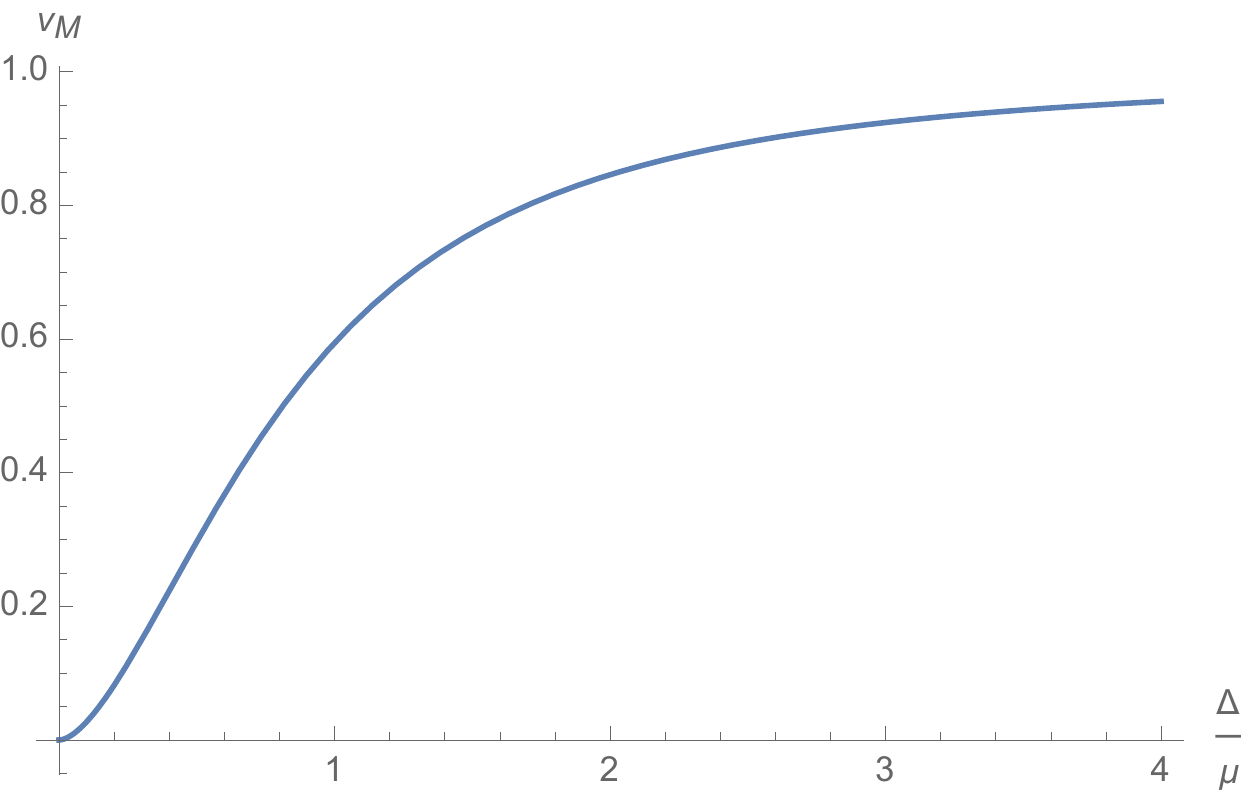}
\caption{The Majorana velocity $v_M$ (in unit of the Fermi velocity $v_F$ ) as a function of the ratio between the SC gap $\Delta$ and chemical potential $\mu$.}
\label{s-alpha}
\end{figure}
Combining the two symmetry requirement above, we found that a general form of the eigenstate at $k_z=0$ is given by
\begin{align}
\chi(r,\theta,k_z=0)=\[f(r)\ g(r)e^{i\theta}\  g^*(r)e^{-i\theta}\ f^*(r)\]^T,
\label{s-wf}
\end{align}
where $r^2=x^2+y^2$. Particle-hole symmetry requires its energy to be zero, and using the fact that $\vec\sigma\cdot {\bf p}=(e^{-i\theta} \sigma^+ + e^{i\theta} \sigma^-)(-i\partial_r)-(e^{-i\theta} \sigma^+ - e^{i\theta} \sigma^-)(1/r)\partial_\theta$,
\begin{align}
&i\partial_r g - \Delta g^* + ig/r  + \mu f = 0 \label{s-5}\\
&i\partial_r g^*+\Delta g  + ig^*/r+ \mu f^* = 0 \label{s-6}\\
&i\partial_r f^* +\Delta f + \mu g^* = 0 \label{s-7} \\
&i\partial_r f -\Delta f^* + \mu g = 0.\label{s-8}
\end{align}
It is easy to check that Eqs.\ (\ref{s-5},\ref{s-6}) and Eqs.\ (\ref{s-7},\ref{s-8}) are consistent only when $g=(1+i)\bar g$ and $f=(1+i)\bar f$. The equations for $\bar f(r)$ and $\bar g(r)$ are 
\begin{align}
&i\partial_r \bar g+i \Delta \bar g + i\bar g/r  + \mu \bar f = 0 \nonumber\\
&i\partial_r \bar f+i\Delta \bar f + \mu \bar g  = 0 \label{s-10}.
\end{align}
Using the ansatz $\bar f(r)=\tilde f(r)\exp(-\int^r \Delta dr')$ and $\bar g(r)=\tilde g(r)\exp(-\int^r \Delta dr')$, we have
\begin{align}
&i\partial_r \tilde g + i\tilde g/r  + \mu \tilde f = 0 \label{s-11}\\
&i\partial_r \tilde f + \mu \tilde g  = 0 \label{s-12}.
\end{align}
Replacing (\ref{s-12}) into \eqref{s-11} we find that $\tilde f$ satisfies the Bessel equation $\partial_r^2 \tilde f + (1/r)\partial_r \tilde f + \mu^2 \tilde f =0$ and $\tilde g$ can be found via Eq.\ (\ref{s-12}). Using the properties of Bessel functions we have in final form
\begin{align}
f(r)=&J_0(\mu r) \exp \[-\int^r \Delta(r') dr'\] (i+1)\nonumber\\
g(r)=&J_1(\mu r) \exp \[-\int^r \Delta(r') dr'\] (i-1).
\end{align}

With the knowledge of the wave function \eqref{s-wf}, we can treat $\delta \mathcal{H}(k_z)=k_z\sigma^z\tau^z$ at a finite but small $k_z$ as perturbation and obtain the small-$k_z$ dispersion. Simple math shows $E(k_z)=v_M k_z$, where the Majorana velocity $v_M$ is given by
\begin{align}
v_M=\frac{\int [J_0^2(r)-J_1^2(r)] \exp(-2\Delta r) dr}{\int [J_0^2(r)+J_1^2(r)] \exp(-2\Delta r) dr},
\end{align}
where we have assumed the the SC gap is a constant (at least away from the vortex).
The integrals are expressed in terms of complete elliptic integrals of first and second kind, $K(x)$ and $E(x)$, as
\begin{align}
v_M=\frac{K(-\frac{\mu^2}{\Delta^2})-\frac{\Delta^2}{\Delta^2+\mu^2} E(-\frac{\mu^2}{\Delta^2})}{E(-\frac{\mu^2}{\Delta^2})-K(-\frac{\mu^2}{\Delta^2})}.
\label{s-377}
\end{align}
We plot $v_M$ as a function of $\Delta/\mu$ in Fig.\ \ref{s-alpha}. We see that in the full range $v_M>0$, which indicates a chiral Majorana mode.

In the limit $\mu=0$ and $\Delta/\mu\to \infty$, we have $f=\exp[-\Delta r](i+1)$ and $g=0$, and the wave function is the Fu-Kane result~\cite{s-Fu-2008} for a superconducting TI surface. In this case one can check that the wave function \eqref{s-wf} is the eigenstate for $\delta \mathcal{H}(k_z)$ (thus the first-order perturbation theory becomes exact), which is consistent with $v_M=1$ (in units where $v_F=1$). On the other hand, in the (more physical) limit where $\Delta\ll \mu$, the Majorana velocity is small but still positive. Expanding Eq.\ \eqref{s-377}, we obtain
\begin{align}
v_M\approx (\Delta/\mu)^2  \log(\mu/\Delta).
\end{align}

For the opposite half vortices $(-1,0)$ and $(0,1)$, it can be straightforwardly verified that the chiral Majorana vortex-core bound state are of opposite chirality. 

\section{IV.~~~Surface states of a 3D $s+ip$ superconductor}

In this section we derive the surface states for a $s+ip$ superconductor, and show that they form a gapped Majorana cone. {A similar derivation can also be found in Ref.\ \onlinecite{s-GR}.}

We assume that the FS is centered around the $\Gamma$ point, and the Bogoliubov-de Gennes (BdG) Hamiltonian expanded around $\Gamma$ point  can be written as
\be
\mathcal{H}=\Psi^\dagger({\bf k})\(-\mu\tau^z + \Delta_p \frac{\bf k}{k_0}\cdot \vec\sigma\tau^x + \Delta_s \tau^y\) \Psi({\bf k}),
\label{s-hsp}
\ee 
where $\mu>0$, $\Psi({\bf k})=\(c({\bf k}),~i\sigma^y c^{\dagger T}(-{\bf k})\)^T$ is the Nambu spinor, $\sigma$ is spin, and $\tau$'s are Pauli matrices in Nambu space. Note that $\Delta_s$ and $\Delta_p$ terms have different Nambu spin structure, indicating their $\pi/2$ phase difference. We have also used the shorthand $\sigma^z\tau^x\equiv \sigma^z\otimes\tau^x$ and $\tau^y\equiv \mathbb{I}\otimes \tau^y$ ($\mathbb{I}$ is a $2\times 2$ identity matrix).

We take open boundary conditions in the $z$-direction, and model the surface as a domain wall of the chemical potential $\mu(z)=\mu\sgn(z)$. Thus, the $z<0$ side is the bulk SC and the $z>0$ side is vacuum. We split the Hamiltonian \eqref{s-hsp} into two parts, $\mathcal{H}_{sp}=\mathcal{H}_1+\mathcal{H}_2$, where
\begin{align}
\mathcal{H}_1=&\Psi^\dagger({\bf k})\[-\mu\tau^z + \frac{\Delta_p}{k_0} ({k_z\sigma^z})\tau^x \] \Psi({\bf k})\nonumber\\
\mathcal{H}_s=&\Psi^\dagger({\bf k})\[ \frac{\Delta_p}{k_0} ({k_x\sigma^x+k_y\sigma^y})\tau^x+\Delta_s \tau^y\] \Psi({\bf k})
\end{align}
The the solution eigenstate of $\mathcal{H}_1$ is standard, and is given by 
\be
\Psi(z)=\Psi_0\exp\(-\frac{\mu k_0}{\Delta_p} \int^z \sgn{z'} dz' \),
\ee
which is a surface bound state at $z=0$, and $\sigma^z\tau^y\Psi_0=\Psi_0$. The eigenvalue of $\mathcal{H}_1$ is zero. Using this wave-function, particular its spinor structure, we find that for wave functions of this type, the effective surface Hamiltonian is
\begin{align}
\mathcal{H}_{\rm surf}=\frac{\Delta_p}{k_0}\(k_x\sigma^y\tau^z-k_y\sigma^x\tau^z\) +\Delta_s\sigma^z,
\end{align}
which is the dispersion of a gapped Majorana cone.

\end{document}